\documentclass[12pt,preprint]{aastex}

\slugcomment{Submitted: ApJ, May 2008}

\shorttitle{TRGB Method}
\shortauthors{Madore et al.}

\begin{document}


\title{Sharpening the Tip of the Red Giant Branch}

\author{\bf Barry ~F. Madore, Violet Mager \& Wendy L. Freedman}
\affil{The Observatories \\ Carnegie Institution of Washington 
\\ 813 Santa Barbara St., Pasadena, CA ~~91101\\}
\email{barry@ociw.edu, vmager@ociw.edu, wendy@ociw.edu\\}




\begin{abstract}
We introduce a modified detection method for measuring the luminosity
of the tip of the red giant branch (TRGB) by introducing the composite
magnitude $T \equiv I - \beta [(V-I)_{\circ} - 1.50]$, where $\beta$
is the slope of the tip magnitude as a function of color (or
metallicity).  The method is specifically designed to account for
known systematics due to metallicity.  In doing so, this simple
transformation does away with arbitrary color selections in measuring
the tip, and thereby significantly boosts the population of resolved
stars that go into defining the TRGB distance.  Moreover this method
coincidentally reduces the impact of reddening on the true modulus as
well as its final uncertainty.
\end{abstract}


 
\section{Introduction}
The luminosity of the tip of the red giant branch (TRGB) has
been profitably used by many groups to determine high-precision
distances to nearby galaxies (for example, see the compilation of
Ferarrese et al.  2001). And, by all measures this method continues to
grow in popularity (e.g., Rizzi et al. 2007 and references therein).
Its success is driven, in all probability, by its low cost in observing
time, its conceptual simplicity, and its wide range of
application (see Madore \& Freedman 1998 for a review).

In our earliest paper on quantifying the procedures for measuring the
magnitude of the TRGB tip (Lee, Freedman \& Madore 1993) we first
recommended the use of a digital (Sobel) filter operating on a binned
histogram representation of the RGB luminosity function. Being
aware of the slight dependence of the RGB tip on metallicity (as
manifest by a monotonic decline of the TRGB I-band magnitude with
increasing color) we also outlined (a regretably convoluted) means of
correcting the observed tip magnitude back to a fiducial
metallicity/color-corrected tip magnitude based largely on
theory. With the clarity of hindsight (drawing on better data, with
continuing theoretical support) we now correct that shortcoming and
suggest a means of measuring a TRGB distance modulus that is
explicitly corrected for metallicity at the filtering level.

In correcting for metallicity the new methodology introduced below is
simple, but its additional utility and power comes from the fact that
it allows the use of a much enhanced sample of tip stars (independent
of their metallicity) to bolster the statistical certainty of the
distance modulus determination.  The method outlined below is, at the
same time, both systematically and statistically more powerful and
robust than previously-used methods.

\section{The $T_{RGB}$ Magnitude}

We begin by introducing the $T$ magnitude for the red giant branch
(RGB). $T_{RGB}$ is the color-corrected $I$-band magnitude
specifically designed to be independent of and insensitive to
metallicity (see Figure 1). By construction

$$ T_{RGB} \equiv I_{\circ} - \beta [(V-I)_{\circ} - \gamma] $$

\noindent
where the color slope $\beta$ is chosen to track the known, and now
calibrated, (metallicity-induced) run of tip magnitude as a function
of intrinsic color, and $\gamma$ is the fiducial color (i.e.,
metallicity) to which the absolute magnitude of the TRGB distance
scale is referred.

The value of $\beta$ can be estimated from theory (see Mager et
al. 2008 for a recent example) and/or derived from observations.  For
a restricted range of intrinsic colors, 1.5 $<$ (V-I)$_{\circ} <$ 3.0
we provisionally adopt a slope of $\beta = $ 0.20 $\pm$ 0.05, which is
consistent with both the value of 0.22 $\pm$ 0.02 given by Rizzi et
al. (2008) for a sample 5 galaxies (NGC~0300, NGC~0598, NGC~1313,
NGC~4258 and NGC~5128), and the value of 0.15 based on theory
(Mager et al.)  The fiducial metallicity, [Fe/H] = -1.7~dex,
corresponds to the bright, low-metallicity end of the RGB which is
marked by stars at the tip having a color of $\gamma$ =
(V-I)$_{\circ}$ = 1.5 mag. The range over which the calibration is
valid extends from here to (V-I)$_{\circ}$ = 3.0~mag, which, for Pop
II stars corresponds to a high metallicity of -0.5 dex.  At the
fiducial color of (V-I)$_o$ = 1.5~mag our absolute magnitude zero
point is set to be M$_{TRGB}$ = -4.05~mag. Thus

$$ \mu_{TRGB} \equiv T_{RGB} - M_{TRGB} =  T_{RGB} + 4.05$$ 

$$ \mu_{TRGB} = I(RGB)_{\circ} - 0.20[(V-I)_{\circ} - 1.50] + 4.05 $$

\noindent
and finally, for the observers:

$$ \mu_{TRGB} = I(RGB)_{\circ}  - 0.20 (V-I)_{\circ} + 4.35 $$

We have noted on previous occasions that, on astrophysical grounds, it
only makes sense for a halo population to contain at least its most
metal-poor stars, given that the low-mass component persists and they
(or rather their originally coeval, high-mass counterparts) are
responsible for the enriched material that may or may not produce
later (higher metallicity) generations. That is, it would be very hard
to have a pure population of high-metallicity Population II stars in the halo
of a galaxy without its low-metallicity (progenitor) component.  Thus
one will always have a low-metallicity sample of stars; and they will
be the first stars detected by the edge filter, because they are also
the brightest TRGB population in any mixture (in the I band).

The real-world limitation comes in having enough stars to fill the
luminosity function at the tip so as not to (downwardly) bias the
detection magnitude.  Simulations addressing this very point (Madore
\& Freedman 1995) suggested that about 100 stars in the magnitude bin
just below the tip are mimimally required to give a credible detection
at the $\pm$0.1~ mag level. As we note below, that may be optimistic.

If one includes only stars of a given color/metallicity when measuring
the tip one then selectively reduces the sample of stars available for
the measurement. On the other hand, by using the metallicity-corrected
sample as advocated here, one can take full advantage of all stars at
the tip, not just those in a narrowly-defined window of color and
metallicity space. If one were to simply open the color range admitted
to the edge-detector without correcting for the color slope one would
certainly increase the sample, but this would clearly be at the
expense of blurring the TRGB discontinuity because of the progressive
bias/contamination by fainter redder (higher-metallicity tip) stars in
the transition region. By seeking a $1/\sqrt{N}$ decrease in the
random error, one would not only be blunting the tool, but also
implicitly inviting an increased systematic error. The method
suggested here corrects for the systematics of metallicity and allows
all available tip stars to contribute to the tip detection.

\section{The New Filtering Methodology}

In this section we discuss details of the application of this
suggested method for TRGB detection and distance determination.

We first need to apply a reddening correction to the observed colors
and apply the appropriate extinction correction to the apparent
magnitudes.  The halos of most galaxies are reasonably assumed to be
dust and extinction free, but any given line of sight may have a
Galactic foreground component. This is usually dealt with by adopting
a predicted value either from the Burstein \& Heiles (1982) or the
Schlegel et al. (1998) Galactic extinction maps.\footnotemark

\footnotetext[1]{It would, of course, be desirable not to have to
depend on foreground dust modelling to determine reddening
corrections, especially if there were some unanticipated dust
component in the host galaxy. Indeed, RGB data alone could be used to
determine a total line-of-sight reddening. This would require adequate
samples of stars below the TRGB to be observed to reasonably good
levels of photometric precision. The method proposed here would equate
the total line-of-sight reddening with the difference in color between
the blue envelope of the observed population of RGB stars compared to
the intrinsic color locus of the most metal-poor calibrating RGB
population. Again, this first population must logically be in the
stellar population mix and it will exclusively populate the blue
edge. More metal-rich, intrinsically redder, stars will only fall in
one direction, away (to the red) from this envelope and will not blur
or bias any reddening determination if the blue envelope is well
defined. Few data sets currently go deep enough below the TRGB, at the
levels of photometric precision required, for this method to be widely
applied to existing observations. That, of course, could change if the
reddening correction were deemed to be critical in any given specific
case.  \ \ \ \ \ An actual implementation of a self-consistent
reddening solution would need to be iterative with the distance
determination itself. The method would boil down to fitting the RGB in
both magnitude and color space by iteratively determining the apparent
magnitude of the tip, and fitting the blue envelope until both the
true modulus and total reddening are self-consistently solved
for. Methods equivalent to the tip detection and its measurement would
have to be deployed to detect and quantitatively measure the blue edge
of the RGB. We defer that implemenation to a future paper, but note,
for the interested reader, that Rizzi et al. (2007) give a detailed
discussion of other methods for independently determining the
reddening which include the red clump, the lower portion of the RGB
and, when available in composite systems, the blue main sequence. }

We next adjust all of the I-band magnitudes using their color
difference with respect to the fiducial color, mentioned above,
$(V-I)_{\circ}$ = 1.50~mag. The fiducial color corresponds to the most
metal-poor and brightest population of TRGB stars expected in our
galaxies. By differentially boosting the I-band magnitudes of the
stars according to their individual colors by -0.20[(V-I)-1.50] mag, we
are in effect forcing all tip stars, regardless of their
metallicities, to take on the same TRBG magnitude.

We can now run the Sobel filter straight down in magnitude space and
be confident that it will directly encounter the full TRGB at right
angles. This guarantees complete and unbiased participation of all tip
stars in the solution; it sharpens the filter output response; and it
eliminates the need for subsequent corrections for mean color or mean
metallicity of the population as a whole.

In practice the data can be binned to a resolution of 0.01~mag without
compromising the precision of the method. This is because such a bin
size exceeds the photometric precision for all but the very brightest
stars in most studies. Typical photometric errors on stars in the data
set to be discussed here are at the TRGB are $\pm$0.03~mag. The basic
filter used supports a traditional [-1, 0, +1] kernel; however, it can
be adjusted to smooth over any number of bins, either to beat down
photometric noise or to decrease the shot noise in sparse
samples. That is, a 3-bin smoothing would effectively employ a [-1,
-1, -1, 0, 0, 0, +1, +1, +1] kernel, smoothing the data and the output
over 0.03~mag intervals. Other kernel weighting schemes are clearly
possible. But the resolution of the tip will never be any better than
the width of the kernel smoothing adopted.

It is useful to estimate the statistical uncertainty in the filter
response to Poisson noise in the binned star counts. Since the filter
response is basically N$_3$ - N$_1$, then the error on that difference
is simply $\pm \sqrt{{N_3}^2 + {N_1}^2}$. The ratio of those two
quantities, (N$_3$ - N$_1$)/$\sqrt{{N_3}^2 + {N_1}^2}$ is a form of
${\chi}^2$, giving not just the output of the filter, but rather the
statistically quantified significance of the output. We plot this
ratio in all of the subsequent figures, and use it as a count of the
number of ``sigmas of significance'' to be associated with any given
response. Furthermore, measuring the semi-width of the response one
unit down from the peak will be taken as the one-sigma uncertainty on
that detection.

We now illustrate the efficacy of this method by examining the various
options as applied to actual data for the halo of NGC~4258.

\section{\bf Experimenting on Real Data}

In the following we make a first pass at illustrating some of the
above claims about the reduced statistical and systematic
uncertainties expected to be associated with the metallicity-corrected
edge detector, as compared to its predecessor(s) which employed standard
filtering at fixed metallicity or color.

For this demonstration we use the ACS data obtained by us for a TRGB
distance determination to the maser galaxy NGC~4258 (Mager et
al. 2008). Over 30,000 stars were in the analysis. These data
provide a nice test case given that the sample size of RGB stars is
large, the photometric errors (at the tip) are small (see below)
and this particular population shows a significant, but not atypical,
one-magnitude spread in (V-I) color.

\subsection{Decreased Sample Size and Bias}

In this section we quantify the effects of focussing the tip detection
algorithm more and more narrowly on the fiducial metallicity (color)
of [Fe/H] = -1.7~dex as represented by stars with colors of (V-I) =
1.5~mag. The filtering algorithm is fixed for all of these runs on the
NGC~4258 dataset and only stars with formal errors of less than
0.30~mag are considered.  Figures 2 through 5 each show  the
color-magnitude diagram in the left panel, and the edge-detection
${\chi}^2$ filter output aligned to the right.  As stated above the
half-width of the filter response one unit down from its maximum will
define our one-sigma figure of merit for the fitting.

We begin the process with no color selection imposed.

In this particular run there were 10,200 stars in the one-mag
(fainter) interval leading up to the TRGB. The tip was measured at I =
25.39 $\pm$ 0.11~mag at a 2.6-sigma significance level (Figure
2). Individual stars at this same magnitude level have typical
photometric errors of $\pm (0.02-0.03)$ mag. In other words, they are
contributing about a third of the quoted variance ($\sigma^2$) for the
tip magnitude. The projected (metallicity-spread induced) magnitude
extent of the TRGB over the color range of 1.5 $<$ (V-I) $<$ 3.0 is
about 0.25~mag, which would have an equivalent sigma of
$\pm$0.08. ``Tip tilt'' is clearly the dominant source of uncertainty
imposed on this measurement of the TRBG magnitude: blurring (and
biasing, see below) the value that would otherwise be obtained at the
fiducial [Fe/H] = -1.7 value alone.

We now investigate the effect of decreased sample size on the
precision and accuracy of the tip detection algorithm. In essence this
is the same test that Madore \& Freedman (1995) ran on
controlled/simulated data; this test is run on real-world photometry
with all of the subtleties and complexities that may or may not be in
any given simulation. In any case, the NGC~4258 data were
progressively restricted in sample size, holding all other parameters
of the tip detection fixed (i.e., maximum error on the photometry of
0.3~mag, and three-bin smoothing of 0.03~mag on the Sobel filter).

Two examples are shown in Figures 4 and 5. The first illustrates a high
significance (1.9 sigma) detection of the tip for a much reduced
sample size, having only about 500 stars in the upper magnitude bin.
The second figure illustrates the effect of reducing the sample
further and the resulting bias in the most significant tip
``detection'' Only about 200 stars were in the magnitude bin below
the (real) tip and the bias is significant.

The summary results for these, and 17 other test samples, are shown in
Figure 6. The individual tip detections are shown as circled
points. The dashed horizontal line shows the asymptotic value of the
tip detection for the full sample of 7,500 stars in the one-magnitude
bin below the TRGB. For sample sizes in excess of 400-500 stars the
solutions agree to within $\pm$0.1~mag (peak-to-peak) which brackets
the commonly quoted statistical uncertainty in this method. However,
for smaller sample sizes the detected tip magnitude begins to
systematically deviate from the fiducial value. The bias should come
as no surprise: as stars drop from the sample there is nowhere for the
tip detector to go but to fainter magnitudes where there are stars and
possibly structure in the RGB luminosity function to trigger on. Of
course the significance of these false tips drops too. Indeed, at
sample sizes below about 400 stars in the upper magnitude bin below
the tip the significance of any ``detection'' starts to drop below the
1.5-sigma level.

In light of these experiments on real data we now revise and extend
our statement concerning the minimum sample sizes needed for TRGB
detection.  For samples having more than 400 stars in the magnitude
bin just below the TRGB the significance of the tip detection should
be expected to be at the 1.5-sigma level or better, with a statistical
uncertainty in the measured tip magnitude being less than
$\pm$0.1~mag. At this level of sample size there are enough stars to
fill the luminsity function up to and including a definition of the
discontinuity.

For sample sizes below this critical value the filling function pulls
away from the discontuity and there is a steep and systematic roll-off
of the (false) tip magnitude toward fainter apparent magnitudes, with
the bias exceeding 0.5~mag at 200-300 stars. Not surprisingly these
false detections will generally be at low ($\sim$one-sigma)
significance levels and should be treated with all due
suspicion. Alternatively, at low count rates one should seriously
consider using maximum-likelihood estimation techniques for measuring
the tip, as discussed by Makarov et al. (2006); they demonstrate
reliable detections when as few as 50 stars are in the one-magnitude
bin below the TRGB.

One final cautionary note: The number of stars below the tip in these
experiments was the number below the known level of the tip. In
practice that value is not known a priori, and so in the regime where
bias sets in that number will be artificially high and the a
posteriori estimate of the bias will be artificially low.

\subsection{Another Form of Bias}

What we have not discussed yet, and what is not possible to address
with the NGC~4258 data, is an opposite bias (toward incorrectly
brighter tip magnitudes). This type of bias may be induced by young
RGB and/or bright AGB stars, especially those resulting from a
superimposed population of disk stars.  Clearly using the $T_{RGB}$
magnitude allows one to optimize the population of stars going into
the tip definition. In practice this means that for a fixed field of
view one can go to fields where the star density is lower than might
have been required for a standard (color-restricted) tip
detection. Moving further out into the halo will reduce any potential
contamination due to disk stars (in those galaxies that have disks),
and this may be of importance in planning future TRGB observations of
composite population systems where the halo is weak and ``disk''
contamination is potentially large. Being able to (radially) distance
one's self from the young and/or intermediate-aged populations will
fast become important in these cases. False signals from AGB
populations in barely resolved galaxies could result in erroneously
small distance moduli, as may well be the case of the reported TRBG
distance to the Antennae galaxies (Saviane et al. 2004), the data for
which were extracted from a Population I rich ``disk'' field
(Schweitzer et al., in preparation)

\subsection{A Small Bonus}

We close out this section with a note about the sensitivity of the
$T_{RGB}$ magnitude to reddening. It is a (happy) coincidence that the
slope of the relation (magnitude versus color) that we are using to
correct TRGB magnitude for metallicity is in the same general
direction as the interstellar extinction/reddening trajectory.  So
when we correct the I-band magnitude by subtracting 0.2~(V-I) we are
also implicitly subtracting 0.2E(V-I) from I. This effectively amounts
to reducing the impact of reddening on the I-band magnitude from A
= 1.4~E(V-I) to A$_T$ = 1.26~E(V-I), with the same 10\% downward
scaling applying to propagating the uncertainty in the extinction.
A small gain, but still in the right direction.

\section{Discussion and Conclusions}

As relatively insensitive to metallicity as the I-band tip of the red
giant branch is, there is still observed to be a slight dependence of
the TRGB magnitude on the metallicity as tracked by the intrinsic
color. We have the means to account for it, since that dependence has been
found to be extremely stable from galaxy to galaxy and is well
calibrated (and indeed well understood). We have incorporated that new
information into a very simply modified TRGB edge detector that makes
distance determinations using the TRGB explicitly independent of the
mean metallicity and/or metallicity distribution of the stars in a
given halo population. In the process this new method allows all
tip stars to add to the statistical significance of the detection of
the core helium ignition discontinuity.  As an added bonus the
modified method is also slightly less sensitive (by 10\%) to reddening
corrections (and/or uncertainties in those same reddening corrections)
as compared to the standard (fixed-metallicity) option.

The method does, of course, now require that both the V and I
magnitudes be obtained for the color-correction to be applied.
Previous applications measuring the tip discontiuity from only an
I-band luminosity function could not be corrected for metallicity
effects, were always met with justifiable skepticism, and are not to be
recommended.

We thank the referee, Brent Tully, for his careful reading of the
paper and his useful suggestions and comments.

\clearpage 

\begin{figure}
\includegraphics [width=14cm, angle=0] {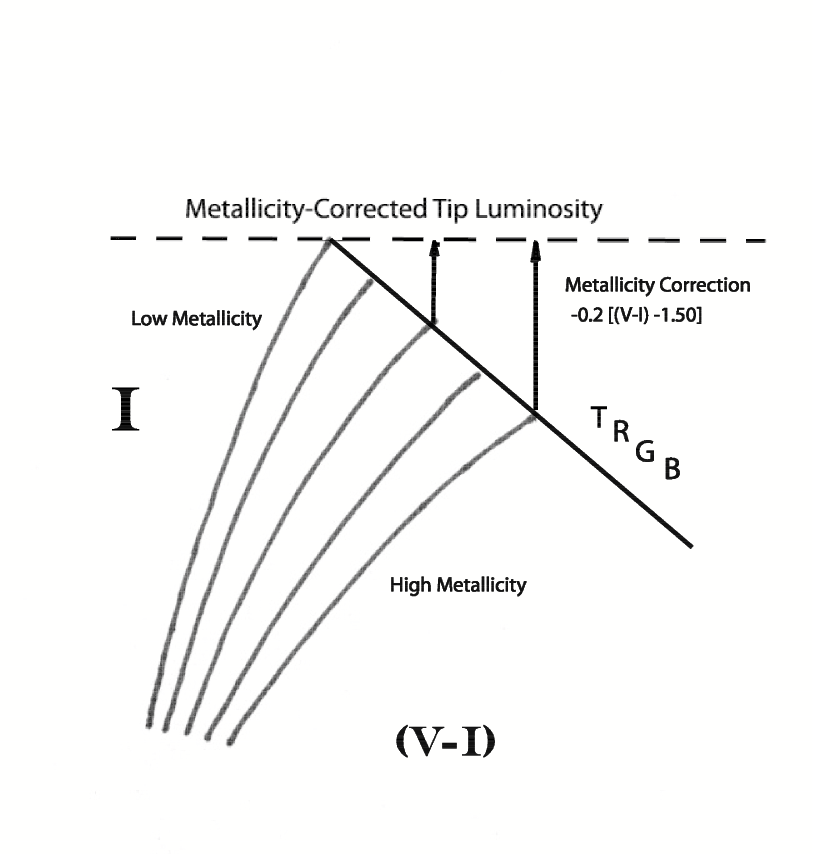} 
\caption{
Schematic representation of a variety
of red giant branches in the I-(V-I) Color-Magnitude Diagram. The
terminal points of the RGB are delineated by the downward sloping line
marked TRGB. Low-metallicity (high-luminosity) giant branches are
shown to the left (blue); high-metallicity giant branches arc up to
the right (red). A color-dependent metallicity correction as applied
to the I-band lumniosity are shown by vertical arrows. These
corrections flatten the TRGB in magnitude space and scale them all to
the lowest-metallicity track.}
\end{figure}

\begin{figure}
\includegraphics [width=18cm, angle=270] {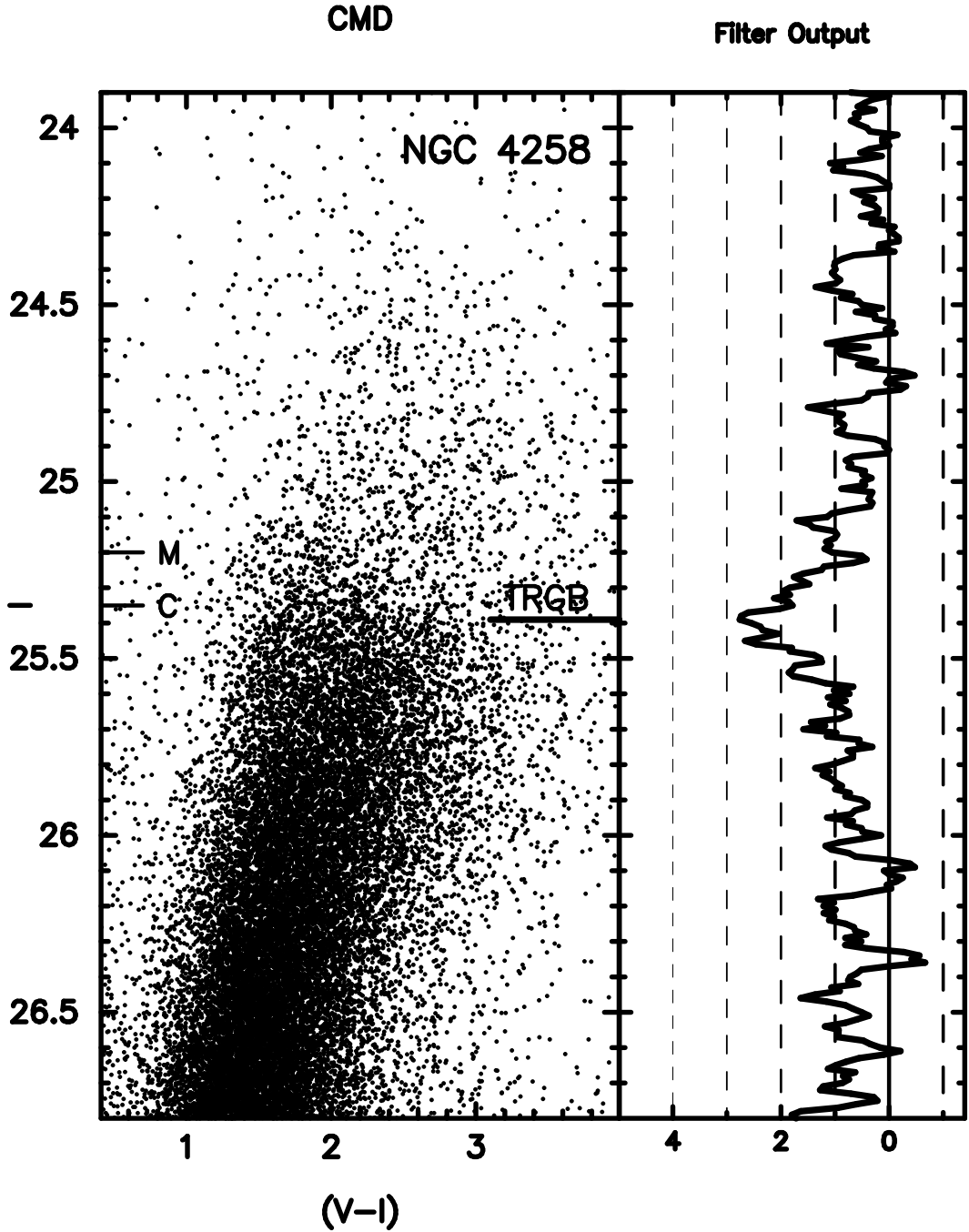} 
\caption{
I vs (V-I) Color-Magnitude Diagram
(CMD) for NGC~4258 to the left, and the normalized Sobel filter
response shown to the right. The tip is measured to be at I = 25.39
$\pm$ 0.11~mag at the 2.6-sigma significance level. M and C mark the
expected magnitude of the tip based on the maser and Cepheid
distances, respectively.}
\end{figure}

\begin{figure}
\includegraphics [width=18cm, angle=270] {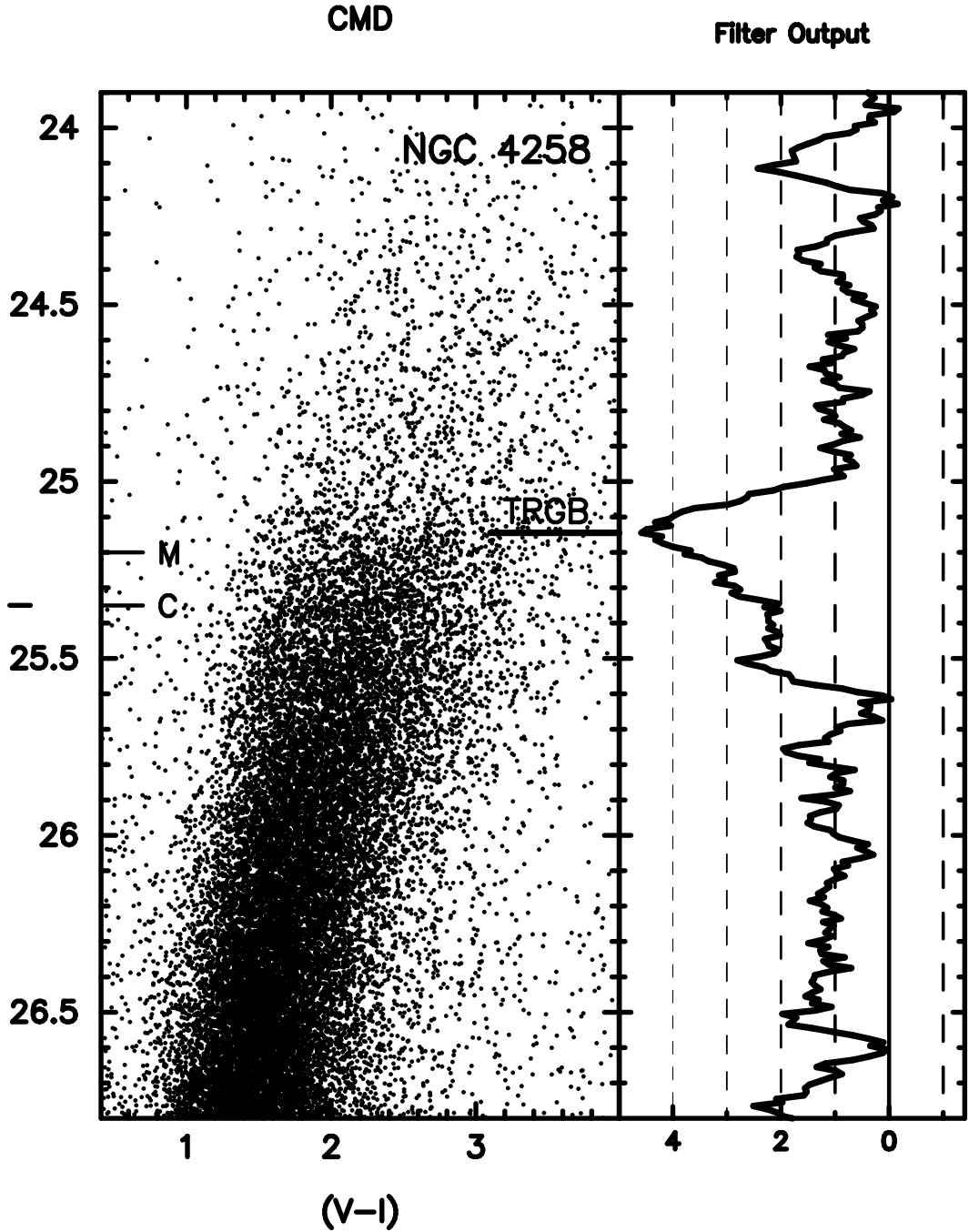} 
\caption{
T$_{RGB}$ vs (V-I)
Metallicity-Corrected Color-Magnitude Diagram for NGC~4258 (see text)
to the left, and the normalized Sobel filter response shown to the
right. The response of highest (relative) significance is marked at I
$= 25.21$~mag with a reported significance of 4.8 sigma.}
\end{figure}

\begin{figure}
\includegraphics [width=18cm, angle=270] {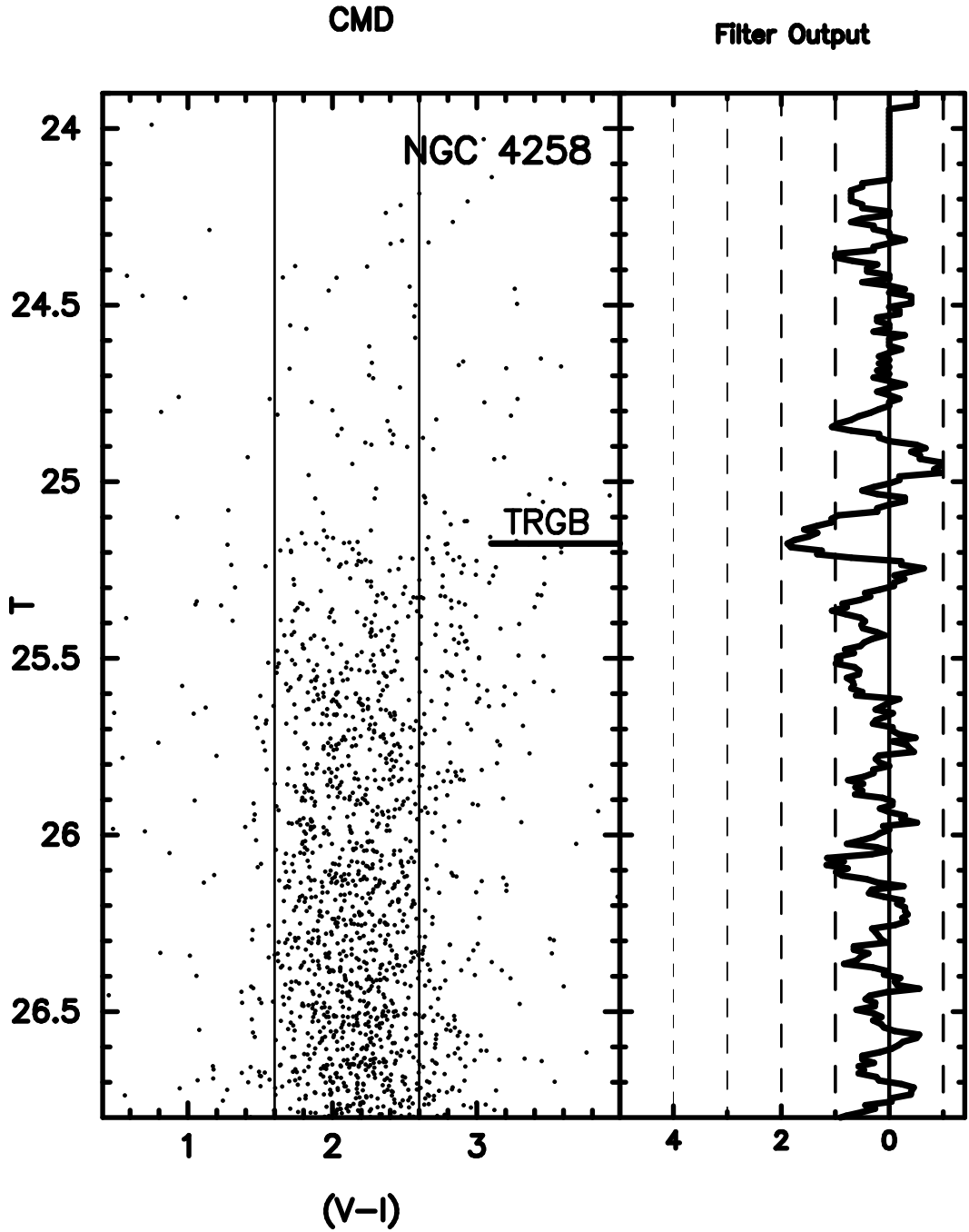} 
\caption{
Same as Figure 3 except for a reduced
sample of NGC~4258 stars. Only one star in 14 was used (amounting to
540 stars in the one-magnitude bin below the tip) for this
tip-detection test. The response of highest (relative) significance is
marked at I $= 25.18$~mag with a reported significance of 1.9 sigma.}
\end{figure}

\begin{figure}
\includegraphics [width=18cm, angle=270] {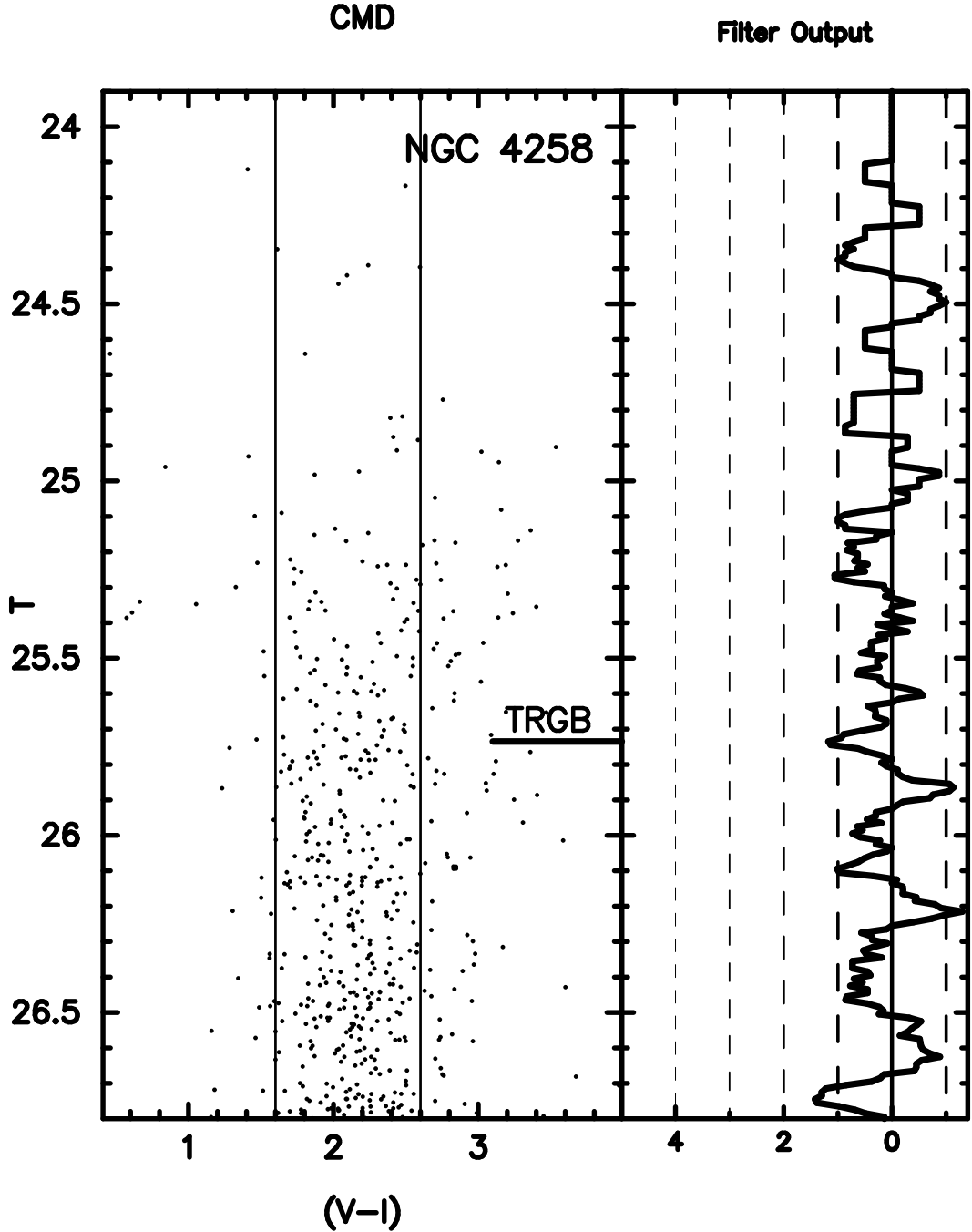} 
\caption{
Same as Figure 3 except for a much reduced sample of
NGC~4258 stars. Only one star in 40 (amounting to only 192 stars in
the one-magnitude bin below the tip) was used for this tip-detection
test. The response of highest (relative) significance is marked at I
$= 25.74$~mag but its absolute significance is low (1.2 sigma) and a
number of other false positives of similar significance are seen
throughout the plot.}
\end{figure}

\begin{figure}
\includegraphics [width=14cm, angle=270] {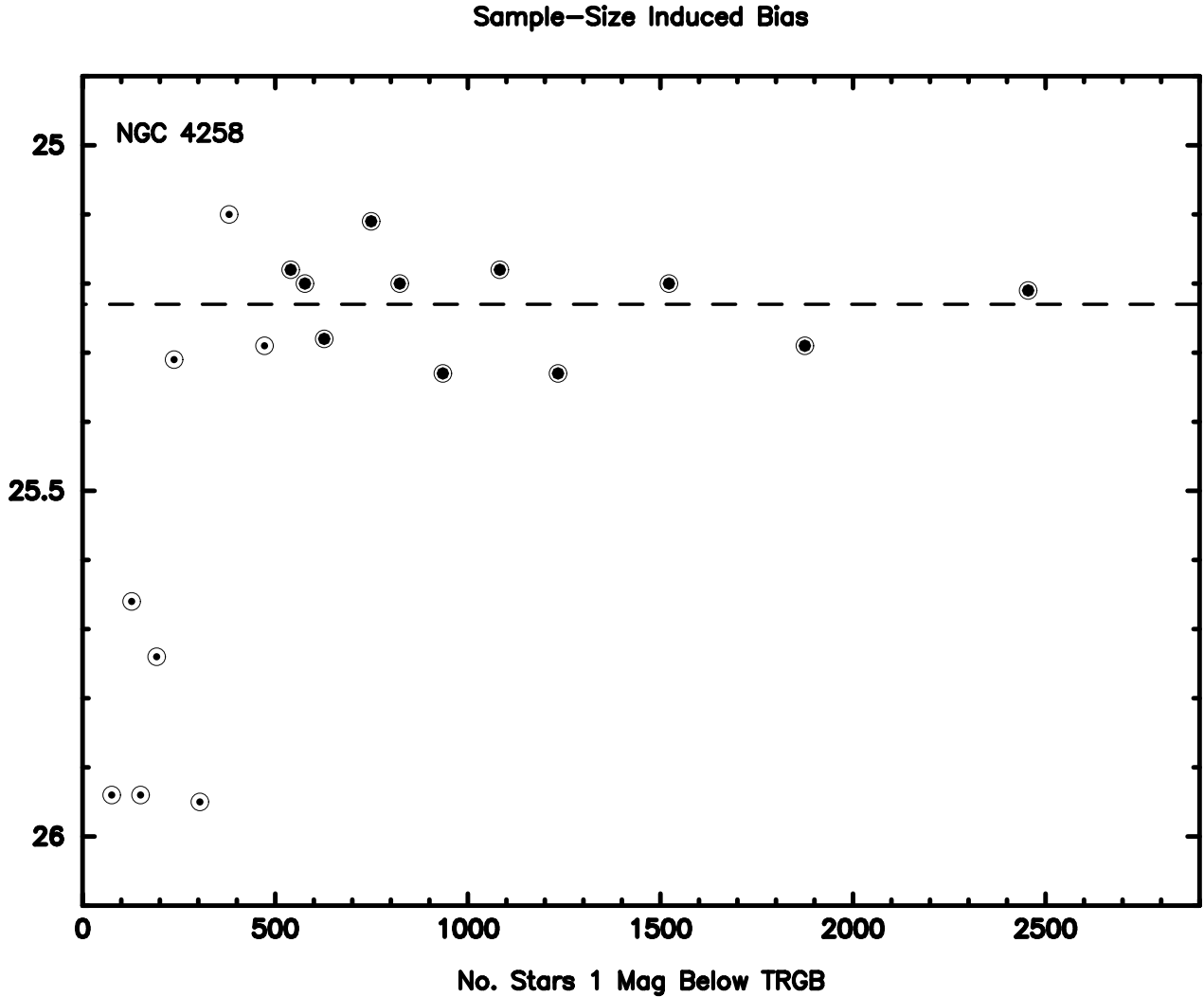} 
\caption{
The run of derived TRGB magnitude with the number of
stars in the magnitude bin below the tip. The dashed line indicates
the true value of the tip magnitude as defined by the largest sample
size available to us (around 7,000 stars in the one-magnitude bin
below the tip). With Poisson noise rising up to about 0.1~mag at the
400-500 star level the tip detections are relatively unbiased for
larger samples. Below that level there is a precipitous fall-off in
the apparent magnitude of the tip resulting in a bias exceeding
0.5~mag for sample sizes less than 200-300 stars in the upper
magnitude bin. Large filled circles have detections that are at, or
above, the 1.5-sigma ($\sim$85\%) confidence level.}
\end{figure}


\begin{references}

\reference{} 
Burstein, D., \& Heiles, C. 1982, AJ, 87, 1165

\reference{} 
Lee, M.G., Freedman, W.L., \& Madore, B.F. 1993, ApJ, 417, 553

\reference{} 
Madore, B.F., \& Freedman, W.L. 1998, ``Stellar
Astrophysics for the Local Group'' VII Canary Islands Winter School,
eds. A. Aparicio, A. Herroro \& F. Sanchez, Cambridge University
Press, Cambridge, p. 234 

\reference{} 
Madore, B.F., \& Freedman, W.L. 1995, ApJ, 109, 1645

\reference{} 
Mager, V., Madore, B.F., \& Freedman, W.L. 2008, ApJ, in press

\reference{} 
Rizzi, L., Tully, R.B., Makarov, D., Makarova, L.,
Dolphin, A.E., Sakai, S., \& Shaya, E.J. 2007, ApJ, 661, 816

\reference{} 
Makarov, D., Makarova, L., Rizzi, L., Tully, R.B., 
Dolphin, A.E., Sakai, S., \& Shaya, E.J. 2006, ApJ, 132, 2729

\reference{} 
Saviane, I., Hibbard, J.E., \& Rich, R.M. 2004, AJ, 127, 660

\reference{} 
Schlegel, D.J., Finkbinder, D.P., \& Davis, M. 1998, ApJ, 500, 525

\end{references}
\end{document}